\documentclass[aps,prl,twocolumn,superscriptaddress,groupedaddress,showpacs]{revtex4}  
\usepackage{graphicx}  
\usepackage{dcolumn}   
\usepackage{bm}        
\usepackage{amssymb,amsmath}   
\usepackage{verbatim}

\usepackage{amsmath,amssymb,amsfonts}

\usepackage{bbold}

\usepackage{float}

\usepackage[usenames,dvipsnames]{xcolor}
\usepackage{epsfig}
\usepackage{epstopdf}
\usepackage{dcolumn}
\usepackage{tikz}
\usetikzlibrary{shapes.geometric, arrows}
\usepackage{upgreek}
\usepackage{setspace}
\usepackage{enumitem}
\usepackage{array,multirow,bigdelim}

\def\be{\begin{equation}}
\def\ee{\end{equation}}
\def\ba{\begin{eqnarray}}
\def\ea{\end{eqnarray}}

\def\a{\alpha}
\def\b{\beta}

\def\b#1{\overline{#1}}

\def\CP1{\mathbb{CP}^1}
\def\SL2C{\mathrm{SL}(2,\mathbb{C})}

\def\Z2{\mathbb{Z}_2}

\def\su2{{SU(2)}}
\def\eps{{\epsilon}}

\def\a{{\alpha}}

\def\[{\left[}
\def\]{\right]}

\def\L{\Lambda}

\def\s{\sigma}
\def\a{\alpha}
\def\b{\beta}

\def\({\left(}
\def\){\right)}
\def\[{\left[}
\def\]{\right]}

\def\<{\langle}
\def\>{\rangle}

\def\i2{\frac{i}{2}}

\def\2F1{\,_2{\rm F}_1}

\newcommand{\beq}{\begin{equation}}
\newcommand{\eeq}{\end{equation}}
\newcommand{\beqq}{\begin{equation*}}
\newcommand{\eeqq}{\end{equation*}}
\newcommand\beqa{\begin{eqnarray}}
\newcommand\eeqa{\end{eqnarray}}
\newcommand\beqaa{\begin{eqnarray*}}
\newcommand\eeqaa{\end{eqnarray*}}
\newcommand\bea{\begin{array}}
\newcommand\eea{\end{array}}
\newcommand{\zz}[2]{z_{#1}-z_{#2}}
\newcommand{\ie}{{\it i.e. }}

\begin{document}

\widetext


\title{New  Factorization Relations for Yang Mills Amplitudes}
\author{N. E. J. Bjerrum-Bohr$^1$,   Poul H. Damgaard$^1$, Humberto Gomez}
\affiliation{ Niels Bohr International Academy and Discovery Center,\\ Niels Bohr Insitute, University of Copenhagen\\
Blegdamsvej 17, DK-2100 Copenhagen \O, Denmark}
\affiliation{ Facultad de Ciencias Basicas,  Universidad Santiago de Cali,\\
Calle 5 $N^\circ$  62-00 Barrio Pampalinda, Cali, Valle, Colombia}

\begin{abstract}
A double-cover extension of the scattering 
equation formalism of Cachazo, He and Yuan (CHY) leads us to conjecture covariant
factorization formulas of  $n$-particle scattering amplitudes 
in Yang-Mills theories. Evidence is given that these factorization
relations are related to Berends-Giele recursions through repeated use
of partial fraction identities involving linearized propagators.

\end{abstract}

\maketitle


\section{Introduction}
The CHY-formalism of scattering equations of Cachazo, He and Yuan provide an intriguing novel way of computing gauge and gravity S-matrix elements~\cite{Cachazo:2013gna,Cachazo:2013hca,Cachazo:2013iea}. The $n$-point scattering amplitudes are here 
expressed in terms of integrals over auxiliary 
variables $z_a$ on the Riemann sphere that become localized on the set of solutions to the scattering equations,\vskip-0.6cm
\be
S_a\equiv \sum_{b=1,b\neq a}^n \frac{s_{ab}}{z_a-z_b}=0\,.
\ee\vskip-0.2cm\noindent
Here $s_{ab} = 2 k_a\cdot k_b$ are generalised Mandelstam variables and the index $a$ labels the (ordered) external particles of 
momenta $k_a$. One remarkable feature of the CHY-formalism, and one which shows its fundamental nature, is that it is
dimension-agnostic. The defining integral over the variables $z_a$ is invariant under an SL(2,${\mathbf C}$)
transformation
\be
z_a ~\to~ \frac{Az_a + B}{Cz_a + D} \,,~~~~~~~~~ AD - BC = 1\,,
\ee
which needs to be fixed. Fixing three of the variables in the standard manner, only $(n-3)$ variables $z_a$ are left. This matches
precisely the $(n-3)$ independent scattering equations after imposing overall momentum conservation. The number of independent
solutions $(n-3)!$ is nevertheless huge and finding all these solutions is computationally difficult even for moderate values of 
$n$. Summing over these independent solutions can fortunately be done more directly, through general integration rules developed
in refs.~\cite{Cachazo:2015nwa,Baadsgaard:2015voa}.  A proof of the CHY formalism has been provided by Dolan and Goddard in ref.~\cite{Dolan:2013isa}. 

Recently, one of us~\cite{Gomez:2016bmv} (see also ref.~\cite{Cardona:2016bpi}) showed how the CHY formalism can be given 
a new formulation in which the basic variables 
$z_a$ live not on $\mathbb{CP}^1$ but on the complex projective plane $\mathbb{CP}^2$. Dubbed the '$\Lambda$-formalism' in~\cite{Gomez:2016bmv}, 
we shall here refer to it as CHY on a double-cover. At first sight it may seem to be a complication
to extend the CHY-formalism in this manner. However, as we shall demonstrate in this paper, the double-cover 
formalism adds a new ingredient to the standard CHY formalism that is much more difficult to extract in the single cover
formulation. Briefly stated, it is this: The 
double-cover formalism naturally expresses the scattering amplitude so that it is factorized into different channels.
The propagator that forms the bridge between two factorized pieces arises as the link between two separate $\mathbb{CP}^1$ pieces, thus
intuitively explaining why the double-cover naturally expresses amplitudes in a factorized manner.

In many cases, the factorizations obtained in this way correspond directly to all the physical channels. Interestingly, there are instances
where unavoidably the factorizations proceed in a slightly different manner: some physical channels appear immediately, but others only
resurface after pole-cancelling terms have rearranged the expressions. 

We start  with a brief review of the CHY-formalism, and then give the corresponding expressions in the double-cover formulation of 
ref.~\cite{Gomez:2016bmv}. Next, we describe how the evaluation of amplitudes on a double-cover produces factorizations into
different channels. Finally we write down an explicit factorization expression valid for $n$ gluons in any dimension and relate
it to known techniques such as on-shell and Berends-Giele recursions.

\section{The CHY Construction and a Double-Cover}

Consider the scattering of $n$ massless particles. The scattering data will then be presented in terms of a set of $n$ momentum vectors 
$\{ k_1^\mu,k_2^\mu,\ldots ,k_n^\mu \}$ and $n$ ``wave functions'' that encode the spin degrees of freedom. For Yang-Mills amplitudes the latter
will correspond to the polarization vectors $\{ \epsilon_1^\mu,\epsilon_2^\mu,\ldots ,\epsilon_n^\mu \}$. Graviton scattering will similarly 
be characterized by a set of polarization tensors, or, simpler, as outer products of polarization vectors. 

%
%

Let us introduce the compact notation of $|ijk|_z$ indicating the Vandermonde determinant of
variables $z_i, z_j, z_k$:
\be
|ijk|_z ~\equiv ~ \prod_{i<j}(z_j - z_i) ~=~  \left|
  \begin{array}{ccc}
    1 & ~z_i & ~z_i^2\\
    1 & ~z_j & ~z_j^2\\
    1 & ~z_k & ~z_k^2\\
  \end{array}
\right|\,.
\ee
It is possible to show that for any rational function $H(z)$ which transforms as
\be\label{int_trans}
H(z)\to H(z)\prod_{a=1}^n(C z_a + D )^4 ,\quad\ee
when
\be \quad z_a \to \frac{A z_a+B}{C z_a+D} \quad{\rm and}\quad AB-CD=1\,,
\ee
the contour integral~\cite{Cachazo:2013hca}
\be
\int \prod_{a=1,a\neq \{i,j,k\}}^n dz_a\, \frac{|ijk|_z |pqr|_z}{\prod_{c=1,c\neq \{p,q,r\}}^n S_c(z)}H(z)\,,
\ee
is independent of the choice of fixed punctures $\{z_i,z_j,z_k\}$ and of equations eliminated $\{S_p,S_q,S_r\}$.

The precise form of the integrand $H(z)$ defines different (color-ordered) theories. 
The simplest case is $\phi^3$-theory. Let us define a 'Parke-Taylor'-factor 
\be
PT(1,2,\ldots,n) ~\equiv ~ \frac{1}{(z_1-z_2)(z_2-z_3)\cdots (z_n-z_1)}\,.
\ee
Color-ordered $\phi^3$-amplitudes correspond to integrands with such factors squared:
\be
H(z) ~=~ \left[PT(1,2,\ldots,n)\right]^2\,.
\ee

As shown in refs.~\cite{Bjerrum-Bohr:2016juj,Bjerrum-Bohr:2016axv} (see also \cite{Cardona:2016gon}), the basic building blocks of other theories
are products of one Parke-Taylor factor with a shuffled Parke-Taylor factor ($\alpha$ indicating a permutation):
\be
H(z) ~=~ PT(1,2,\ldots,n)\times PT(\alpha(1),\alpha(2),\ldots,\alpha(n))\,.
\ee
Such a product of Parke-Taylor factors in the integrand thus forms a basic skeleton for all other theories. 

For Yang-Mills theory we have
\be
H^{\rm YM}_n  = 
{ PT}{(1,2,...,n)}
\times {\rm Pf}^\prime \Psi_n, \ee
where   
\be
{\rm Pf}^\prime \Psi_n \equiv \frac{(-1)^{i+j}}{\zz{i}{j}}\,\,{\rm Pf}[(\Psi_n)^{ij}_{ij}].\ee
The  $2n\times 2n$ matrix, $\Psi_n$, is defined as 
\vspace{-0.1cm}
\begin{eqnarray}\label{Pmatrix}
\Psi_n  \equiv  \left( 
\begin{matrix}
 A & -C^{\rm T} \\
C  & B 
\end{matrix}
\right)\,\, , 
\end{eqnarray}\vskip-0.4cm\noindent
with, 
\be
A_{ab} \equiv \begin{cases} \displaystyle 
\frac{\a^2 \, k_{a}\cdot k_b}{\zz{a}{b}} \\
\displaystyle \quad ~~ 0  \end{cases} \!\!\!\!\!\!\!, \ \ \ \ \ B_{ab} \equiv \begin{cases} \displaystyle \frac{\eps_a\cdot \eps_b}{\zz{a}{b}} & a\neq b\,,\\
\displaystyle \quad ~~ 0 & a=b\,,\end{cases}
\ee
and\vskip-0.7cm
\be\label{C-matrix}
C_{ab} \equiv   \begin{cases} \displaystyle \frac{\a\,\eps_{a}\cdot k_b}{\zz{a}{b}} & a\neq b\,,\\
\displaystyle  -\sum_{c=1;c\neq a}^n \frac{\a\,\epsilon_a \cdot k_c}{ \zz{a}{c}}  & a=b\,.\end{cases}
\ee\vskip-0.2cm\noindent
Notice the unusual normalization in the $A$ and $C$ matrices. If we put $\alpha=1$ we recover the CHY-prescription as originally defined. If instead we choose $\a=\sqrt{2}$ the normalization matches with the color-ordered Feynman rules given by Dixon in \cite{Dixon:1996wi}. In what follows, $\alpha$ can
take any value (it only changes the overall normalization of the color-ordered amplitudes, a convention), but we keep it arbitrary at this point to facilitate
a comparison with Feynman diagrams based on color-ordered Feynman rules later in this paper.
The matrix $(\Psi_n)^{ij}_{ij}$ denotes the reduced matrix obtained by removing the rows and columns $i, j$ from $\Psi_n$, where $1 \leq i< j \leq n$. For how to use the integration rules \cite{Baadsgaard:2015voa,Baadsgaard:2015ifa} in the context of Yang-Mills theory, see \cite{Bjerrum-Bohr:2016juj, Cardona:2016gon, 
Bjerrum-Bohr:2016axv}.

\subsection{The Double-Cover}

A double-cover version of the CHY construction was recently developed by one of us in~\cite{Gomez:2016bmv}. 
In this approach the amplitudes are 
given as contour integrals on $n$-punctured double-covered Riemann spheres. Restricted to the curves
$0 = \mathsf{C}_a \equiv y_a^2 - \sigma_a^2 + \Lambda^2$ for $a = 1,\ldots,n$, the pairs $(\sigma_1,y_1), (\sigma_2,y_2),\ldots, (\sigma_n,y_n)$
provide the new set of doubled variables. A translation table has been worked out in detail in ref.~\cite{Gomez:2016bmv}.
Specifically, one defines
\begin{equation}
\tau_{(a,b)}\equiv \frac{1}{2(\s_{a}-\s_{b})}\left( \frac{y_a + y_b+ \s_{a}-\s_{b}}{y_a}\right),
\end{equation}
and 
\begin{equation}
         \Delta_{(pqr)} \equiv  \big( \tau_{(p,q)}\,\tau_{(q,r)}\,\tau_{(r,p)}\big)^{-1} \,                 
\end{equation} 
and simultaneously imposes scattering equations in the form (momentum conservation $\sum k_a=0$ is implicitly used throughout)
\begin{equation}
S_a^{\tau}\equiv 
\sum_{b=1 \atop b\neq a}^n s_{ab} \,\tau_{(a,b)}=0\,,
\end{equation}
where $a=1,\ldots , n$.
Amplitudes are then derived from the following expression:
\begin{equation}\label{Lprescription}
{A}_n^\Lambda=\int_{\Gamma} d\mu^{\L}_n\times\frac{{\cal I}_n(\sigma,y)}{ S^{ \tau}_m}\,,
\end{equation}
where the measure $d\mu_n^{\L}$ is defined as
\begin{equation}\label{measure}
d\mu_n^{\L}\equiv \frac{1}{{\rm Vol}({\rm GL}(2,\mathbb{C}))} \times \frac{d\L}{\L}\,\prod_{a=1}^n 
\frac{y_a\, dy_a\,d\s_a}{\mathsf{C}_a}\, \frac{\Delta_{(pqr)} }{\prod_{d\neq p,q,r,m} S^{\tau}_d}\,,
\end{equation}
with the  $\Gamma$  contour being defined by the equations
\!\!\begin{equation}
\begin{cases}\L=0 \\ S^{\tau}_d=0\end{cases}\!\!\!{\rm for}~ d\neq \{p,q,r,m\}, ~\mathsf{C}_{1}=0,\ldots, \mathsf{C}_{ n} =0.
\end{equation}
This rewriting of the amplitude in terms of this contour $\Gamma$, which does
not encircle the scattering equation $S^{\tau}_m$ follows from the Global Residue Theorem.
Note that the integrand now includes a scale $\Lambda$.
In order to fix this larger ${\rm GL}(2,\mathbb{C})$ symmetry, we gauge-fix four $\s_a$'s. Then the measure must be multiplied  
by the Faddeev-Popov determinant
\begin{eqnarray}
&&
\hspace{-0.47cm}
\Delta_{ (pqr|m) } \equiv 
\s_p\Delta_{ (qrm) }
- \s_m \Delta_{(pqr)} +\s_r  \Delta_{(mpq)} - \s_q  \Delta_{(rmp)}. \nonumber\\     
&&                    
\end{eqnarray}
Therefore,  $d\mu_n^{\L}$ becomes
\begin{equation}\label{measureGF}
d\mu_n^{\L}=
\frac{1}{2^2}
\frac{d\L}{\L} \, \prod_{a=1}^n \frac{y_a\, dy_a}{\mathsf{C}_a} \, \!\!\!\!\!  \prod_{d\neq p,q,r,m}\!\!\frac{d\s_d}{ S^{\tau}_d}\times\Delta_{(pqr|m)}\, \,\Delta_{(pqr)}\,,
\end{equation}
which has been explained in detail in \cite{Gomez:2016bmv,Gomez:2018cqg}.

As in the original CHY approach, the precise form of the integrand ${\cal I}_n(\s,y)$ defines the theory. 
For example, color-ordered $\phi^3$-theory corresponds to the integrand
\begin{equation}\label{malphabeta}
{\cal I}_n ~=~  [   PT^\tau(1,2,\ldots n)]^2 \,,
\end{equation}
where
\begin{equation}
 PT^\tau(1,2,\ldots n) \equiv \tau_{({1,2})}\,\tau_{(2,3)}\cdots \tau_{(n,1)} \,.
\end{equation} 
Note the $\tau$'s are neither antisymmetric nor symmetric; the precise definition as given above is correct.
Similarly, other theories correspond to products of such modified Parke-Taylor factors with additional expressions, much like in the original
CHY formalism. Again, the integrands for these other theories can be broken down to products of shuffled Parke-Taylor expressions.%

\section{The  Yang-Mills Theory in the Double-Cover Prescription}\label{sectionPfaffian}

Since $\tau_{(a,b)}\neq - \tau_{(b,a)}$, it is not immediately obvious how to define the double-cover analog of the reduced Pfaffian for  pure Yang-Mills theory.
In order to obtain the double-cover version of the $\Psi_n$ matrix, we write (we define $(y\s)_a\equiv y_a+\s_a$)
\begin{equation}
\hspace{-0.1cm}
\tau_{(a,b)} =\frac{(y\s)_a}{y_a} \times \, T_{ab} \equiv \frac{(y\s)_a}{y_a} \times \, 
\frac{1}{(y\s)_a-(y\s)_b}, 
\end{equation}
on the support, $\mathsf{C}_a=\mathsf{C}_b=0$, where clearly, $T_{ab}= -T_{ba}$.
Since $T_{ab}$ is anti-symmetric, we establish the single and double-cover identification, $\displaystyle\frac{1}{\s_{ab}} \, \leftrightarrow \, T_{ab}$,  so, the double-cover matrix, $\Psi^\L_n$, is defined as, $\Psi^{\L}_n\equiv \Psi_n\Big|_{\frac{1}{\s_{ab}}\, \rightarrow \, T_{ab}}$.
Notice that it is straightforward to rewrite the $\phi^3$-integrand in terms of $T_{ab}$, namely
\begin{equation}
{\cal I}^{\phi^3}_n(\a|\b) =  { PT^\tau}{(\a_1,\ldots \a_n)} \times 
 \prod_{a=1}^n    \frac{(y\s)_a}{y_a}  \times
{ PT}^T{(\b_1,\ldots \b_n)}  ,  \nonumber
\end{equation}
with,
\be
{ PT}^T{(\b_1,\b_2,\ldots \b_n)} \equiv T_{\b_1\b_2}\, T_{\b_2 \b_3 }\cdots T_{\b_n \b_1} \,.
\ee

Following the CHY program developed in \cite{Cachazo:2013iea}, the double-cover representation of the ordered Yang-Mills amplitude is obtained by the replacing,  ${  PT}^T{(\b_1,\b_2,\ldots \b_n)} \rightarrow (-1)^{i+j}\,T_{ij}\,{\rm Pf}[(\Psi^\L_n)^{ij}_{ij}]$, \ie   
\begin{equation}
{\cal I}^{\rm YM}_n(\a) ~=~   PT^\tau{(\a_1,\ldots \a_n)} \times 
{\rm Pf}^{\prime} \Psi^\L_n \,,
\end{equation}
with
\begin{equation}\label{Pfdef}
{\rm Pf}^{\prime} \Psi_{n}^\L \equiv  \prod_{a=1}^n   \frac{(y\s)_a}{y_a}   
\times
(-1)^{i+j}\,T_{ij}\,\,
 {\rm Pf} \left[(\Psi^\L_n)^{ij}_{ij}\right]\, ,
\end{equation}
where the $(\Psi^\L_n)^{ij}_{ij}$ matrix is given by removing the rows and columns $i, j$ from $\Psi^\L_n$, with $1 \leq i< j \leq n$.
Therefore, the pure Yang-Mills amplitude at tree-level  in the double-cover language is given by the expression
\begin{equation}\label{YMgeneric}
A_n(\a) = \int_\Gamma d\mu^\L_n \, 
 \frac{(-1)\,\Delta_{(pqr)} \, \Delta_{(pqr|m)}} {S^\tau_m } \times
{\cal I}_n^{\rm YM}(\a), 
\end{equation}
where the upper index ``${\rm YM}$" in $A_n(\a) $ is no longer necessary.

\section{A Simple Example}

As a simple example, let us consider the four-point amplitude, $A_4(1,2,3,4)$, with the gauge fixing $(pqr|m)=(123|4)$ and the reduced matrix, $(\Psi^\L_4)^{13}_{13}$. 

First, we focus on the configuration where the sets of punctures $(\sigma_1,\sigma_2)$ and $(\sigma_3,\sigma_4)$ are respectably on the upper and the lower sheet of the curves \\[-15pt]
\begin{align}
(y_1=+\sqrt{\s_1^2-\L^2},\s_1), \quad (y_2=+\sqrt{\s_2^2-\L^2},\s_2),  \\[-3pt]
(y_3=-\sqrt{\s_3^2-\L^2},\s_3),  \quad (y_4=-\sqrt{\s_4^2-\L^2},\s_4). \nonumber
\end{align} \\[-10pt]
Expanding  all elements in $A_4(1,2,3,4)$ around $\L=0$,  we obtain (to leading order)
\begin{align}
	 PT^\tau(1,2,3,4)\Big|^{1,2}_{3,4}= \frac{\L^2}{2^2} \frac{1}{(\s_{12} \s_{2P_{34}} \s_{P_{34}1} )} 
	\frac{1} { (\s_{P_{12}3} \s_{34} \s_{4P_{12}} )}, \nonumber
\end{align}
\begin{align}\label{TTeq}
&\frac{\Delta(123) \Delta(123|4)}{ S^{\tau}_4}
\Big|^{1,2}_{3, 4}  
= \\&\hskip0.8cm  \frac{2^5}{\L^4} (\s_{12}\, \s_{2P_{34}} \,  \s_{P_{34}1}  )^2 
 \left( \frac{ 1 }{ P^2_{34}} 
\right) 
\times (\s_{P_{12}3} \, \s_{34} \, \s_{4P_{12}})^2,
 \nonumber 
\end{align}
\vspace{-0.3cm}
\begin{eqnarray}
&&
\hspace{-0.5cm}
 \prod_{a=1}^4   \frac{(y\s)_a}{y_a}   
\times T_{13} \,
{\rm Pf}\left[
(\Psi_4^\L)^{13}_{13}
\right]\Big|^{1,2}_{3,4} = -  \frac{  \L^2  }{2^2 } \times \nonumber \\
&&
\hspace{-0.5cm}
\sum_M  \,
 \frac{ (-1) }{\s_{P_{34}1 }} \times
{\rm Pf}\left[
{\small
\begin{matrix}
0 & - \a\frac{\eps^M_{{34}}\cdot k_{2}}{\s_{P_{34}2}} & -  \a\frac{\eps_{1}\cdot k_{2}}{\s_{12}} & - {C}_{22}\\
 \a \frac{\eps^M_{{34}} \cdot k_{2}}{\s_{P_{34}2}} & 0 & \frac{\eps^M_{{34}} \cdot \eps_{1}}{\s_{P_{34}1}} & \frac{\eps^M_{{34}} \cdot \eps_{2}}{\s_{P_{34}2}}\\
 \a  \frac{\eps_{1}\cdot k_{2}}{\s_{12}} &  \frac{\eps_{1}\cdot \eps^M_{{34}}  }{\s_{1P_{34}}} &  0 & \frac{\eps_{1}\cdot \eps_{2}}{\s_{12}} \\
{ C}_{22} &   \frac{\eps_{2}\cdot \eps^M_{{34}}   }{\s_{2P_{34}}} &  \frac{\eps_{2}\cdot \eps_{1}}{\s_{21}} &  0  \\
\end{matrix}}
\right] \times
 \nonumber \\
&&
\hspace{-0.5cm}
 \frac{ (-1) }{\s_{P_{12}3 }} \times
{\rm Pf}\left[
{\small
\begin{matrix}
0 & -  \a \frac{\eps^M_{{12}}\cdot k_{4}}{\s_{P_{12}4}} & - \a  \frac{\eps_{3}\cdot k_{4}}{\s_{34}} & - { C}_{44}\\
 \a \frac{\eps^M_{{12}}\cdot k_{4}}{\s_{P_{12}4}} & 0 & \frac{\eps^M_{{12}}\cdot \eps_{3}}{\s_{P_{12}3}} & \frac{\eps^M_{{12}}\cdot \eps_{4}}{\s_{P_{12}4}}\\
 \a  \frac{\eps_{3}\cdot k_{4}}{\s_{34}} &  \frac{\eps_{3}\cdot \eps^M_{{12}}}{\s_{3P_{12}}} &  0 & \frac{\eps_{3}\cdot \eps_{4}}{\s_{34}} \\
{ C}_{44} &   \frac{\eps_{4}\cdot \eps^M_{{12}}}{\s_{4P_{12}}} &  \frac{\eps_{4}\cdot \eps_{3}}{\s_{43}} &  0  \\
\end{matrix}}
\right]   \label{pfexpansion}\\
&&
\hspace{-0.5cm}
=-  \frac{  \L^2  }{2^2 } \, \sum_M
\frac{ (-1) }{\s_{P_{34}1 }}\,
{\rm Pf}\left[
\left(\Psi_3 \right)^{P_{34}1}_{P_{34}1}
\right] \, 
\times
 \frac{ (-1) }{\s_{P_{12}3 }}\,
{\rm Pf}\left[
\left( \Psi_3 \right)^{P_{12}3}_{P_{12}3}
\right], \nonumber
\end{eqnarray}
where we have introduced the notation, $P_{ij} \equiv  k_i +k_{j}$, and the
new fixed punctures, $\s_{P_{34}}=\s_{P_{12}}=0$. The $C_{22}$ and $C_{44}$ factors are given by the usual expressions   \footnote{It is useful to recall that, $\eps_{2}\cdot P_{34} = -\eps_{2}\cdot k_{1}$ and $\eps_{4}\cdot P_{12} = -\eps_{4}\cdot k_{3}$.}, ${C}_{22} =- \a \frac{\eps_2\cdot k_1}{\s_{21}} - \a \frac{\eps_2\cdot P_{34}}{\s_{2P_{34}}} $, ${C}_{44} =- \a \frac{\eps_4\cdot k_3}{\s_{43}} - \a \frac{\eps_4\cdot P_{12}} {\s_{4P_{12}}} $, and the equality in \eqref{pfexpansion} is obtained under the completeness relationship
\begin{align}
	\label{eq:epsM}
	\sum_M\epsilon_{i}^{M\, \mu} \epsilon_{j}^{M\, \nu} = \eta^{\mu\nu}.
\end{align}
Therefore,  the labels sets, $\{1,2\}$ and $\{3,4\}$, have been separated. 

From the measure, $d\mu_4^\L = \frac{1}{2^2}\frac{d\L}{\L}$, we compute the $\L$ integral and the amplitude becomes
\begin{align}
& 
A_4(1,2,3,4) \Big|^{1,2}_{3,4} =  \\
&
\frac{1}{2} \sum_M  \frac{A_3(P^{\epsilon^M}_{34},1,2)
\times A_3(P^{\epsilon^M}_{12},3,4)  }{P^2_{12}}   ,
\quad\nonumber
\end{align}
\noindent 
where on the right hand side the factorized object is given by amplitudes with one leg off-shell, as indicated.  
The overall factor $1/2$ cancels out after summing over mirrored configurations, {\it i.e.}, 
\begin{align}\label{schannel}
&
A_4(1,2,3,4) \Big|^{1,2}_{3,4} +A_4(1,2,3,4) \Big|_{1,2}^{3,4} \nonumber \\
&
=  \sum_M  \frac{A_3(P^{\epsilon^M}_{34},1,2)
\times A_3(P^{\epsilon^M}_{12},3,4)  }{P_{12}^2} .
\end{align}
\noindent 

In a similar way, the factorization expansion,  $A_4(1,2,3,4) \Big|^{4,1}_{2,3} $, becomes 
\begin{align}\label{tchannel}
&
A_4(1,2,3,4) \Big|^{4,1}_{2,3} +A_4(1,2,3,4) \Big|^{2,3}_{4,1} \nonumber \\
&
=  \sum_M  \frac{A_3(P^{\epsilon^M}_{41},2,3)
\times A_3(P^{\epsilon^M}_{23},4,1)  }{P_{23}^2} .
\end{align}
\noindent 

Notice that after starting with the double-cover reduced matrix, $(\Psi_4^\L)^{ij}_{ij}=(\Psi_4^\L)^{13}_{13}$, the resulting subamplitudes  in \eqref{schannel} and \eqref{tchannel} have as reduced matrices the ones  obtained by removing the rows/columns
{\small
\begin{eqnarray}\label{removingij}
\{ i,j \}=\{ {\rm off}\text{-} {\rm shell\,\, puncture} \} \cup  \left( \{\rm All\,\, punctures \} \cap \{1,3 \} \right)\!,~~
\end{eqnarray}
}
as it can be seen  in \eqref{pfexpansion}.

Finally, besides to the two physical factorization expansions around $\L=0$   achieved previously, from the double-cover approach arises a spurious channel given by  $A_4(1,2,3,4) \Big|^{1,3}_{2,4} $, up to its mirrored configuration. At leading order, this configuration is expanded  as

\begin{align}
	 PT^\tau(1,2,3,4)\Big|^{1,3}_{2,4}= \frac{\L^4}{2^4} \frac{1}{(\s_{1P_{24}} ^2\, \s_{3P_{24}}^2 )} 
	\frac{1} { (\s_{2P_{13}}^2\,   \s_{4P_{13}}^2 )}, \nonumber
\end{align}
\begin{align}\label{TTeq2}
&\frac{\Delta(123) \Delta(123|4)}{ S^{\tau}_4}
\Big|^{1,3}_{2, 4}  
= \\&\hskip0.8cm  \frac{2^5}{\L^4} (\s_{13}\, \s_{3P_{24}} \,  \s_{P_{24}1}  )^2 
 \left( \frac{ 1 }{ P^2_{13}} 
\right) 
\, (\s_{P_{13}2} \, \s_{24} \, \s_{4P_{13}})^2,
 \nonumber 
\end{align}
\vspace{-0.3cm}
\begin{eqnarray}
&&
\hspace{-0.5cm}
 \prod_{a=1}^4   \frac{(y\s)_a}{y_a}   
\,\, T_{13} \,
{\rm Pf}\left[
(\Psi_4^\L)^{13}_{13}
\right]\Big|^{1,3}_{2,4} \!\!= -  \frac{\a^2\, P_{13}^2}{2} \frac{(\eps_1\cdot \eps_3) \, (\eps_2\cdot \eps_4)   }{\s_{13}^2 \s_{24}^2} \nonumber \\
&&
\hspace{-0.5cm}
=\frac{ ( \s_{1P_{24}} \s_{3P_{24}}  ) ( \s_{2P_{13}} \s_{4P_{13}}  )  }{\s_{13} \, \s_{24}} \times 2 \times
\nonumber
\\
&&
\hspace{-0.5cm}
\left\{ 
\sum_L  \,
 \frac{ (-1) }{\s_{P_{24}1 }} \times
{\rm Pf}\left[
{\small
\begin{matrix}
0 & - \a\frac{\eps^L_{{24}}\cdot k_{3}}{\s_{P_{24}3}} & -  \a\frac{\eps_{1}\cdot k_{3}}{\s_{13}} & - {C}_{33}\\
 \a \frac{\eps^L_{{24}} \cdot k_{3}}{\s_{P_{24}3}} & 0 & \frac{\eps^L_{{24}} \cdot \eps_{1}}{\s_{P_{24}1}} & \frac{\eps^L_{{24}} \cdot \eps_{3}}{\s_{P_{24}3}}\\
 \a  \frac{\eps_{1}\cdot k_{3}}{\s_{13}} &  \frac{\eps_{1}\cdot \eps^L_{{24}}  }{\s_{1P_{24}}} &  0 & \frac{\eps_{1}\cdot \eps_{3}}{\s_{13}} \\
{ C}_{33} &   \frac{\eps_{3}\cdot \eps^L_{{24}}   }{\s_{3P_{24}}} &  \frac{\eps_{3}\cdot \eps_{1}}{\s_{31}} &  0  \\
\end{matrix}}
\right] \times \right.
 \nonumber \\
&&
\hspace{-0.5cm}
\left.
 \frac{ (-1) }{\s_{P_{13}2 }} \times
{\rm Pf}\left[
{\small
\begin{matrix}
0 & -  \a \frac{\eps^L_{{13}}\cdot k_{4}}{\s_{P_{13}4}} & - \a  \frac{\eps_{2}\cdot k_{4}}{\s_{24}} & - { C}_{44}\\
 \a \frac{\eps^L_{{13}}\cdot k_{4}}{\s_{P_{13}4}} & 0 & \frac{\eps^L_{{13}}\cdot \eps_{2}}{\s_{P_{13}2}} & \frac{\eps^L_{{13}}\cdot \eps_{4}}{\s_{P_{13}4}}\\
 \a  \frac{\eps_{2}\cdot k_{4}}{\s_{24}} &  \frac{\eps_{2}\cdot \eps^L_{{13}}}{\s_{2P_{13}}} &  0 & \frac{\eps_{2}\cdot \eps_{4}}{\s_{24}} \\
{ C}_{44} &   \frac{\eps_{4}\cdot \eps^L_{{13}}}{\s_{4P_{13}}} &  \frac{\eps_{4}\cdot \eps_{2}}{\s_{42}} &  0  \\
\end{matrix}}
\right] 
\right\}
  \label{pfexpansion2}\\
&&
\hspace{-0.5cm}
=
\frac{ ( \s_{1P_{24}} \s_{3P_{24}}  ) ( \s_{2P_{13}} \s_{4P_{13}}  )  }{\s_{13} \, \s_{24}} \times  \nonumber
\\
&&
\hspace{-0.2cm}
2 \times
\sum_L
\frac{ (-1) }{\s_{P_{24}1 }}\,
{\rm Pf}\left[
\left(\Psi_3 \right)^{P_{24}1}_{P_{24}1}
\right] \, 
\times
 \frac{ (-1) }{\s_{P_{13}2 }}\,
{\rm Pf}\left[
\left( \Psi_3 \right)^{P_{13}2}_{P_{13}2}
\right], \nonumber
\end{eqnarray}
with, $\s_{P_{13}} = \s_{P_{24}}=0$, and  $\sum_{L}$ means a sum over longitudinal degree of freedoms, namely
\begin{align}
	\label{eq:epsL}
	\sum_L\epsilon_{i}^{L\, \mu} \epsilon_{j}^{L\, \nu} = \frac{P_i^\mu\, P_j^\nu}{P_i \cdot P_j } .
\end{align}
Considering the above expansions we are able to 
integrate the measure, $d\mu_4^\L = \frac{1}{2^2}\frac{d\L}{\L}$, so, it is straightforward to see 
\begin{align}\label{sp-channel}
&
A_4(1,2,3,4) \Big|^{1,3}_{2,4} +A_4(1,2,3,4) \Big|_{2,4}^{1,3} \nonumber \\
&
=\left. - 2 \times \sum_L  \frac{A_3(P^{\epsilon^L}_{34},1,2)
\times A_3(P^{\epsilon^L}_{12},3,4)  }{P_{12}^2} \right|_{2 \leftrightarrow 3} .
\end{align}
\noindent 
Therefore, the double-cover approach give us the four-point factorization relation

\begin{align}\label{f-pFactorized}
&
A_4(1,2,3,4) =   \sum_M  \frac{A_3(P^{\epsilon^M}_{41},2,3)
\times A_3(P^{\epsilon^M}_{23},4,1)  }{P_{23}^2}    \nonumber \\
&
+\sum_M  \frac{A_3(P^{\epsilon^M}_{34},1,2)
\times A_3(P^{\epsilon^M}_{12},3,4)  }{P_{12}^2}
\nonumber \\
&
\left. - 2 \times \sum_L  \frac{A_3(P^{\epsilon^L}_{34},1,2)
\times A_3(P^{\epsilon^L}_{12},3,4)  }{P_{12}^2} \right|_{2 \leftrightarrow 3} ,
\end{align}
\noindent 
where the subamplitudes  are given in the single-cover approach with reduced matrices satisfying the eq.
\eqref{removingij}.

\section{A new relation for Yang-Mills amplitudes}

We now wish to generalize the new  factorization realization obtained from double-cover formalism in the previous section. As will be shown in great detail elsewhere ~\cite{Gomez:2018cqg}, by integrating the double-cover representation of an ordered Yang-Mills amplitude one is led to the following general formula which factorizes arbitrary $n$-point Yang-Mills aplitudes into a product of (single-cover) CHY representations of lower-point amplitudes: \vskip-0.4cm
\begin{eqnarray}\label{Gen-C}
&& 
\hspace{-0.4cm}
A_n \big( 1,\ldots , n\big)   =  \sum_{\epsilon_{\!M}} \frac{A_{3}\big({{P}^{\epsilon_{\!M}}_{\!4:1}},2,3\big) \!\times \!A_{n-1}\big({{P}^{\epsilon_{\!M}}_{\!2:3}},4,\ldots, n,1\big) }{ P_{23}^2}  \nonumber \\
&& \hspace{-0.4cm}+
\! \sum_{i=4}^{n}\!\! \left[ \sum_{\epsilon_{\!M}}\!
\frac{A_{n-i+3}\big({P}^{\epsilon_{\!M}}_{\!3:i},i\!+\!1,\dots 1,2\big)\, A_{i-1}\big({{{P}^{\epsilon_{\!M}}_{\!i+1:2 }}},{3},\dots,i \big) }{P_{i+1:2}^2} \right. \nonumber \\
&&
\hspace{-0.2cm}
\!\!\!\!\! - 2\! \!\left. \left. \sum_{\epsilon_{\!L}}\!
\frac{A_{n-i+3}\big({P}^{\epsilon_{\!L}}_{\!3:i},i\!+\!1,\dots 1,2\big)\, A_{i-1}\big({{{P}^{\epsilon_{\!L}}_{\!i+1:2 }}},{3},\dots,i \big) }{P_{i+1:2}^2}\right|_{2\leftrightarrow 3} \right]
\nonumber\\
\end{eqnarray}
Let us be clear: this factorized form of Yang-Mills amplitudes is a conjecture. What the double-cover formalism produces directly are the two first terms
plus contributions that come from linking amplitudes together with scalar degrees of freedom. Miraculously, it appears that these scalar
contributions can be exactly represented by gluing two Yang-Mills amplitudes together with longitudinal polarizations only. The technical details of how
these manipulations arise will be presented elsewhere \cite{Gomez:2018cqg}. Needless to say, in the factorized form on the right hand side the two 
amplitudes each have one external leg off-shell (although still dressed with the corresponding unphysical polarization vector). Gluing these two 
amplitudes together proceeds through the polarization sums as described in the eqs.  \eqref{eq:epsM} and \eqref{eq:epsL}.
It should also be stressed that the above expression comes from the double-cover formalism with Mobius and scale-invariance gauge choices $(pqr|m)=(123|4)$ and reduced matrix $(\Psi_n^\L)^{13}_{13}$. 

This is important to remark 
since the above factorization is a gauge-fixing dependent expression. Of course, the final result, the left hand side, is the correct full $n$-point amplitude, but the precise factorized form on the right hand side depends on that generalized gauge fixing. The three punctures which must be fixed in the smaller off-shell Yang-Mills amplitudes are given by the set, $\{ \rm fixed\,\, punctures \} = \left( \{\rm All\,\, punctures \} \cap \{1,2,3,4 \} \right) \cup\{\rm off \text{-}shell\,\, puncture\}$, and their reduced matrices are obtained by removing the rows/columns under the rule given in \eqref{removingij}. We denote sums of cyclically-consecutive external momenta (modulo the total number of external momenta) by $P_{i: j} \equiv  k_i +k_{i+1} +\ldots+k_{j-1} +k_j$.  For expressions with only two momenta involved (not necessarily consecutive) we are using the shorthand notation $P_{ij} \equiv k_i+k_j.$  

We have denoted the  polarization degrees of freedom by $\epsilon_{\!M}$ and longitudinal by $\epsilon_{\!L}$. 
Using the simple identity $\sum_{\!M} \eps_{i}^{M\,\mu} \eps_{j}^{M\,\nu} \; = \sum_{\!T} \eps_{i}^{T\,\mu} \eps_{j}^{T\,\nu} \;  + \sum_{\!L} \eps_{i}^{L\,\mu} \eps_{j}^{L\,\nu} \; $, where,  $\sum_{\!T} \eps_{i}^{T\,\mu} \eps_{j}^{T\,\nu} \;=\eta^{\mu\nu}\;- \frac{ P_{i}^{\mu}  P_{ j}^{\nu} }{ P_{i}\cdot  P_{ j} }$, we can rewrite \eqref{Gen-C} in terms of transverse (T) and longitudinal (L) polarization vectors,  
\begin{eqnarray}\label{Gen-CT}
&& 
\hspace{-0.4cm}
A_n \big( 1,\ldots , n\big) =\sum_{\epsilon_{\!T}} \frac{A_{3}\big({{P}^{\epsilon_{\!T}}_{\!4:1}},2,3\big) \,  A_{n-1}\big({{P}^{\epsilon_{\!T}}_{\!2:3}},4,\ldots, n,1\big) }{ P_{23}^2}     \nonumber  \\
&& \hspace{-0.4cm}
+
\! \sum_{i=4,\epsilon_{\!T}}^{n}\!\! \frac{A_{n-i+3}\big({P}^{\epsilon_{\!T}}_{\!3:i},i\!+\!1,\dots 1,2\big) \, A_{i-1}\big({{{P}^{\epsilon_{\!T}}_{\!i+1:2 }}},{3},\dots,i \big) }{P_{i+1:2}^2}  \nonumber \\
&&
\hspace{-0.4cm}
 - 2 \!\! \!\nonumber 
 \left.
\sum_{i=4,\epsilon_{\!L}}^{n}\!\! \frac{A_{n-i+3}\big({P}^{\epsilon_{\!L}}_{\!3:i},i\!+\!1,\dots 1,2\big) \, A_{i-1}\big({{{P}^{\epsilon_{\!L}}_{\!i+1:2 }}},{3},\dots,i \big) }{P_{i+1:2}^2}  \right|_{2\leftrightarrow 3}
\\
&& 
\hspace{-0.4cm}
+\sum_{i=4,\epsilon_{\!L}}^{n} \!\! 
\frac{A_{n-i+3}\big({P}^{\epsilon_{\!L}}_{\!3:i},i\!+\!1,\dots 1,2\big)\, A_{i-1}\big({{{P}^{\epsilon_{\!L}}_{\!i+1:2 }}},{3},\dots,i \big) }{  P_{i+1:2}^2} \nonumber \\
&& 
\hspace{-0.4cm} 
+\! \sum_{\epsilon_{\!L}} \frac{A_{3}\big({{P}^{\epsilon_{\!L}}_{\!4:1}},2,3\big) \, A_{n-1}\big({{P}^{\epsilon_{\!L}}_{\!2:3}},4,\ldots, n,1\big) }{P_{23}^2} 
\end{eqnarray}

Notice that the poles related to the longitudinal polarization contributions are not physical and indeed these unphysical poles are
cancelled by corresponding numerator factors. This is the way local 4-point Yang-Mills interactions appear in this formalism.

\subsection{ Feynman diagrams and Bern-Carrasco-Johansson (BCJ) numerators}
We will first consider how the double-cover representation relates to BCJ numerator identities \cite{Bern:2008qj}. From the formula \eqref{Gen-CT}, we arrive at
\begin{eqnarray}\label{fourP-Fey}
&&\!\!\!\!\!\!\!\!\!\!\!\!\!\!\!\!\!\!\!A_4 (1,2, 3,4)\nonumber
=\\&&\quad\quad
\sum_{\epsilon_T}\frac{   A_{3}(P_{\!\!12}^{\epsilon_T} \!, {3} , 4)\!\times\!  A_{3} (P^{\epsilon_T}_{\!\!34}\!,{1} , 2)  }{P_{\!\!12}^2} \nonumber \\
&&\quad\quad+\sum_{\epsilon_T}\frac{  A_{3} (P_{\!\!23}^{\epsilon_T}\!, 4,1) \!\times\!  A_{3} (P_{\!\!41}^{\epsilon_T}\!,2,3)   } {P_{\!\!41}^2}
\nonumber\\
&&\ \ - 2 \sum_{\epsilon_{\!L}}  \Big[ \left.
\frac{   A_{3}(P_{\!\!12}^{\epsilon_L} \!, {3} , 4)\!\times\!  A_{3} (P^{\epsilon_L}_{\!\!34}\!,{1} , 2)  }{P_{\!\!12}^2} \right|_{2 \leftrightarrow 3}
\nonumber\\ 
 &&\quad\quad\quad - \frac{  A_{3} (P_{\!\!12}^{\epsilon_{\!L}} , {3} , 4)\times
 A_{3} ({ P_{\!\!34}^{\epsilon_{\!L}}},{1} , 2)  }{2P_{\!\!12}^2} 
\nonumber 
\noindent \\ &&\quad\quad\quad - \frac{ A_{3}  ({P_{\!23}}^{\epsilon_{\!L}}, 4,{1} ) \times  A_{3} ({P_{\!\! 41}^{\epsilon_{\!L}}},2,{ 3} )   }{2P_{\!\! 41}^2} 
\Big]\,.
\end{eqnarray}

\vskip-0.35cm\noindent
It is simple to check that in the normalization convention $\a=\sqrt{2}$ (corresponding to \cite{Dixon:1996wi}),  the first and second line are just 
the conventionally normalized Feynman diagrams,  $\parbox[c]{2.4em}{\includegraphics[scale=0.062]{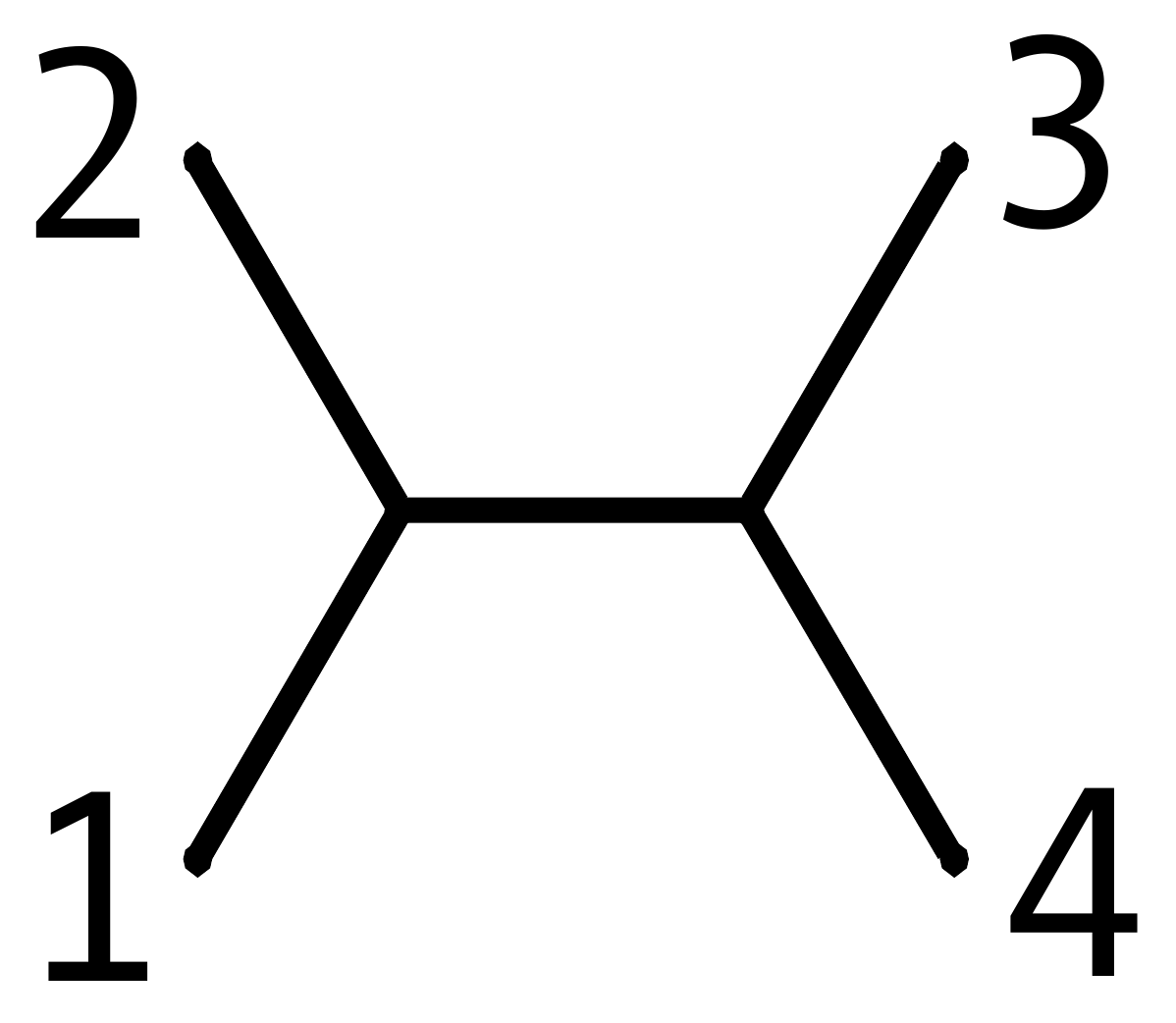}} $ and $\parbox[c]{2.6em}{\includegraphics[scale=0.062]{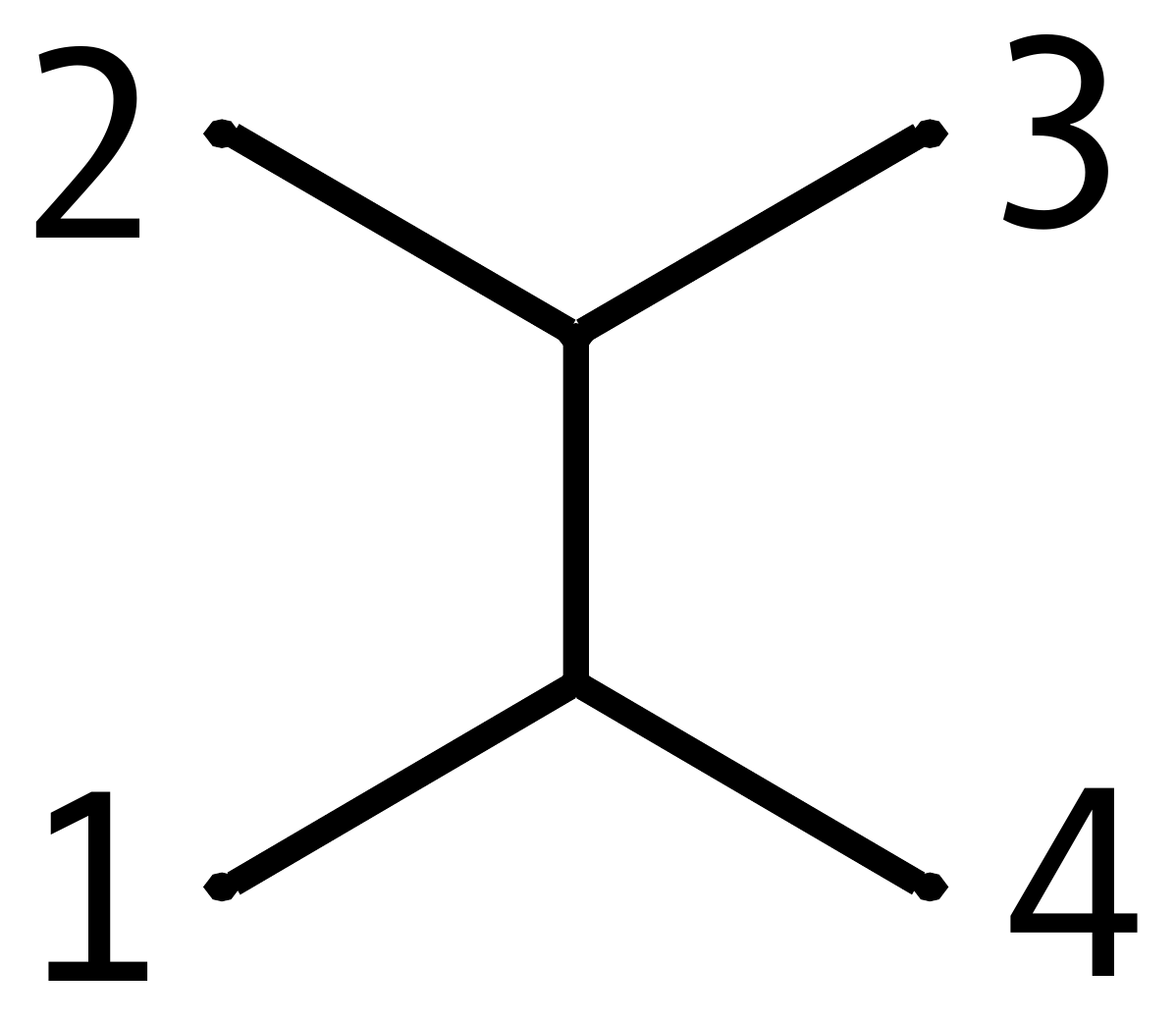}}$ and the remainder represents the quartic vertex, namely $\parbox[c]{2.6em}{\includegraphics[scale=0.062]{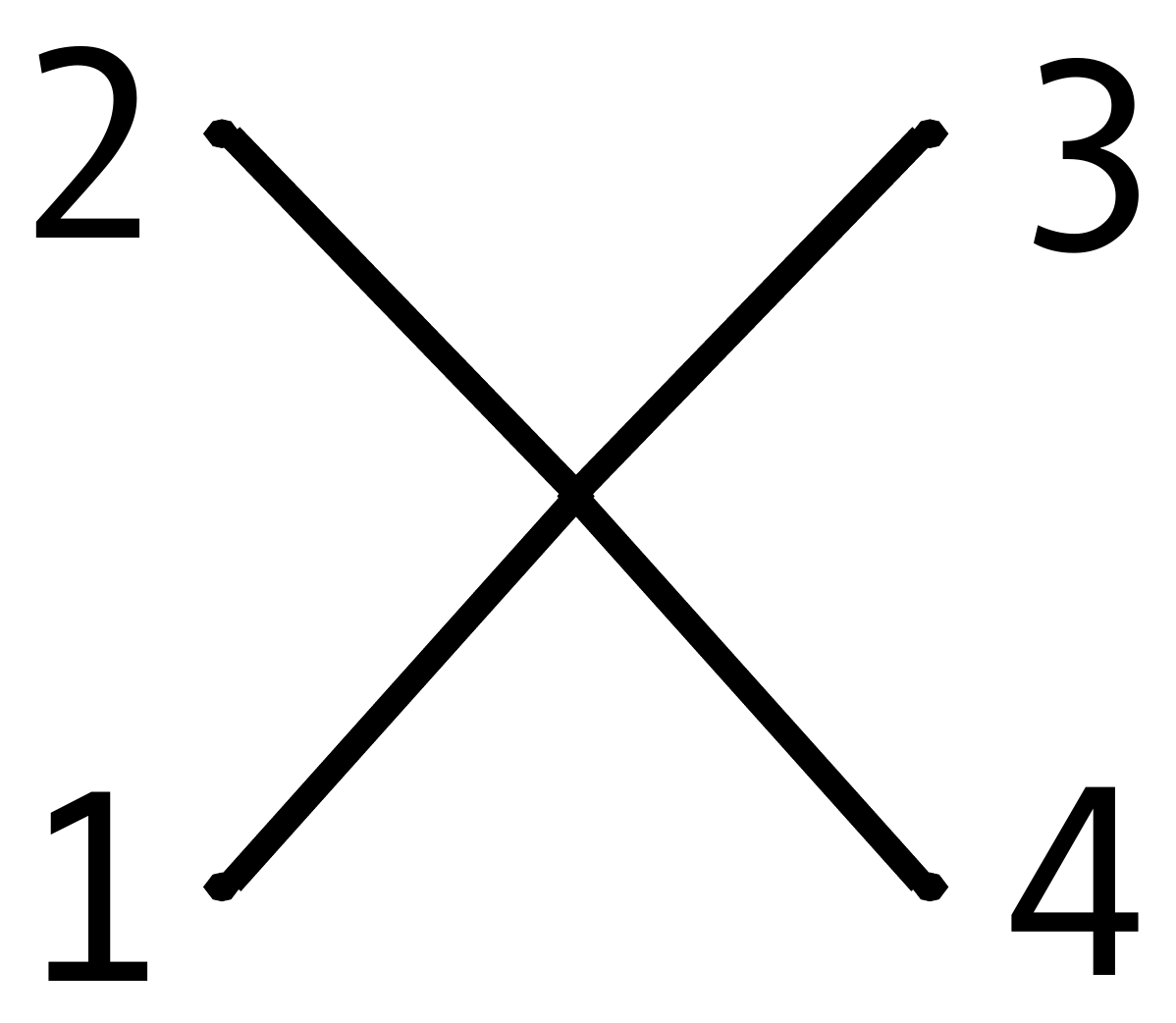}}$.  Finally, to obtain the BCJ numerators, we reorganize \eqref{fourP-Fey} in the following way
\begin{eqnarray}\label{}
\hspace{-2.6cm}
A_4 (1,2, 3,4)=\frac{{\bf n}_{s}  }{P_{\!\! 12}^2} + \frac{{\bf n}_{t}  }{P_{\!\! 41}^2}, \quad {\rm with},
\nonumber
\end{eqnarray}\vskip-0.5cm
\begin{eqnarray}\label{}
\hspace{-0.6cm}
{\bf n}_{s}
=
 \sum_T  A_{3} (P_{\!\! 12}^{\epsilon_T} , 3 , 4)\times A_{3} (P_{\!34}^{\epsilon_T},1,2) + 
 \nonumber \\
 \sum_{\!L}
P_{\! 12}^2\Big[ - \left.
 \frac{A_{3} (P_{\!\! 12}^{\epsilon_L} , 3 , 4)\times A_{3} (P_{\!34}^{\epsilon_L},1,2) }  {P_{\!\! 12}^2} \right|_{2 \leftrightarrow 3}
\\ +
\frac{ A_{3}^{(1)}(P_{\!23}^{\epsilon_{\!L}}, 4,1)  \times A_{3}^{(3)}(P_{\!41}^{\epsilon_{\!L}},2,3)  }{P_{\!\! 41}^2}
\Big],
\nonumber
\end{eqnarray}\vskip-0.4cm
\begin{eqnarray}
{\bf n}_{t}
=
 \sum_T  A_{3} (P_{\!23}^{\epsilon_T} , 4,1)\times  A_{3}  ( P_{\!41}^{\epsilon_T}, 2,3)+\nonumber\\
\sum_{\!L}
P^2_{\!14} \Big[
- \left.
 \frac{A_{3} (P_{\!\! 12}^{\epsilon_L} , 3 , 4)\times A_{3} (P_{\!34}^{\epsilon_L},1,2) }  {P_{\!\! 12}^2} \right|_{2 \leftrightarrow 3}
\\
 +
\frac{ A_{3}^{(3)}( P_{\!\!12}^{\epsilon_{\!L}}, 3, 4)\times A_{3}^{(1)}(P_{\!34}^{\epsilon_{\!L}}, 1, 2)   }{P_{\!\!12}^2}
\Big]\,.
\nonumber
\end{eqnarray}
Using the above it is simple to check that we have,     
${\bf n}_{s} - {\bf n}_{t} = {\bf n}_{u}$, where ${\bf n}_{u}$ can be obtained from ${\bf n}_{s}$ under the permutation, $(1,2,3,4) \rightarrow (1,3,2,4)$. Extending such ideas to a higher number of points should be a possible avenue and would be very interesting.

\subsection{BCFW recursion}
It is interesting to compare the factorizations above with what one would obtain based 
on Britto-Cachazo-Feng-Witten (BCFW) recursion \cite{Britto:2005fq}). To illustrate, consider the five-point amplitude, $A^{}_{5}(1,2,3,4,5)$, and introduce the momentum deformation
\begin{equation}
k^\mu_2(z) = k^\mu_2 + z \,q^\mu, \quad k^\mu_3(z) = k^\mu_3 - z \,q^\mu, \quad z\in\mathbb{C},
\end{equation}
where $q^\mu$ satisfies, $k_2\cdot q=k_3\cdot q=q\cdot q=0$ and $q\cdot \bar q=1$. Additionally, the polarization vectors, $\{\eps_2,\eps_3\}$, must be deformed in order to keep the transversality, so we consider, $\eps^+_2(z) =\bar q - z\frac{k_3}{k_2\cdot k_3}$ and $\eps^+_3(z) =q $, or another option is, 
$\eps^-_2(z) =q $ and $\eps^-_3(z) =\bar q + z\frac{k_2}{k_2\cdot k_3}$.
Since we have momentum conservation for deformed momenta  $k_1+k_2(z)+k_3(z)+k_4+k_5=0$ and the on-shell condition $k^2_2(z)=k^2_3(z)=0$ and transversality remain valid, the CHY approach is well defined. Thus, 
from \eqref{Gen-C} and using Cauchy one has
\begin{eqnarray}
&&
\hspace{-0.5cm}
A_5 ({1},2,{3},4,5)= 
\nonumber \\
&&
\hspace{-0.5cm}
  - {\rm Res}_{ P^2_{\!34}(z) =0} \Big[ \frac{\sum_{\!M}  A_3 ( P_{\!5:2}^{\epsilon_M},  3, 4)(z) \!\times\!  A_4  (P_{\! 34}^{\epsilon_M}, 5,1,2)(z) }{z \, P_{\! 34}^2(z)} 
\Big]  
   \nonumber \\
&&
\hspace{-0.5cm}
  - {\rm Res}_{ P^2_{\!3:5}(z) =0} 
\Big[  
\frac{\sum_{\!M} A_3 (P_{\!3:5}^{\epsilon_M}, 1,2)(z)\, A_4 ( P_{\!12}^{\epsilon_M} ,3, 4, 5) (z) } {z\, P_{\!3:5}(z) }
\Big]
 \nonumber \\
 &&
 \hspace{-0.5cm}
  - {\rm Res}_{z=\infty} 
 \left[
 \frac{A_5 (1,2,3,4,5) (z) }{z}
 \right].
\nonumber
\end{eqnarray}
Obviously the pole $P_{23}^2$ does not depend on $z$ so that this physical factorization channel only contributes at infinity. The most interesting 
observation is that the spurious poles, $P^2_{\!\!i{+}1:2}(z)\Big|_{2 \leftrightarrow 3}$ cancel out because the longitudinal contributions  $\sum_{i=4,\epsilon_{\!L}}^{n} \!\!   A_{n-i+3} \big({{P_{\!\!3:i}^{\epsilon_{\!L}}}},i\!+\!1,\ldots,1,2\big)\!\times\! A_{i-1}\big({P_{\!\!i{+}1:2}^{\epsilon_{\!L}}}, 3,4, ... ,i \big)\Big|_{2 \leftrightarrow 3} $ are proportional to them. Therefore,
 the boundary contributions at $z=\infty$ are related to the unphysical poles that appear in the double-cover, eq. \eqref{Gen-C}. This gives these poles a special significance in the context of BCFW recursion and potentially a new recursive path for dealing with such contributions.

\subsection{ Berends-Giele recursion and the double-cover}
Another natural question arises concerns the similarity of the factorized forms from the double cover double-cover method and Berends-Giele recursion  \cite{Berends:1987me}. In order to shed light on this we will will focus in the bi-adjoint $\phi^3$ theory in the double-cover formalism.  Because
of the trivial numerator factors of this case it is far simpler to analyze.

The connection is well illustrated by considering the five-point amplitude. The factorizations from the double-cover method lead to
\vspace{-0.1cm}
{\small
\begin{eqnarray}\label{Lalgo}
&&
\hspace{-0.6cm}
 A^{\phi^3}(1,2,3,4,5) = \frac{ A_4^{\phi^3} (P_{\!12},3,4,5) } {P_{\!3:5}^2}
+ \frac{ A_4^{\phi^3} (1,P_{\!23},4,5) } { P_{\!4:1}^2} \nonumber  \\
&& \hspace{3.3cm}
+ \frac{ A_4^{\phi^3} (1,2,P_{\!34},5) } { P_{\!34}^2}   \\
&& 
\hspace{0.1cm}
= \frac{1}{P_{\!3:5}^2} \!\left(    \frac{1}{ P_{\!5:2}^2\!-\!P^2_{\!12} } \!+\! \frac{1}{P_{\!45}^2 } \right)  \!+\!
 \frac{1}{P^2_{\!34} } \!\left(    \frac{1}{ P^2_{\!51} }  \!+\! \frac{1}{  P^2_{\!3:5}\! -\!P^2_{\!34}  } \right)   
 \nonumber \\
&& 
\hspace{1.3cm}\nonumber
+
 \frac{1}{P^2_{\!4:1}} \left(    \frac{1}{ P^2_{51} } \!+\! \frac{1}{  P_{\!45}^2 } \right) \,,
\end{eqnarray}
}
\vskip-0.1cm\noindent
where we have chosen the gauge fixing, $(pqr|m)=(123|4)$.
On the other hand, Berends-Giele recursion gives (see, e.g., ref. \cite{Mafra:2016ltu})
\vspace{-0.1cm}
{\small
\begin{eqnarray}\label{mafraF}
\frac{1}{ P_{\!2:4}^2}\left(  \frac{1}{P_{\!34}^2} + \frac{1}{P_{\!23}^2} \right)  
+\frac{1}{P_{\!12}^2 \, P_{\!34}^2} 
+ \frac{1}{P_{1:3}^2} \left(  \frac{1}{P_{12}^2} + \frac{1}{P_{23}^2} \right) \,\,.
\end{eqnarray}
}
\vskip-0.1cm\noindent
On the support, $k_1+k_2+k_3+k_4 +k_5=0$,  and under the on-shell condition, $k_i^2=0$, it is trivial to check that the expressions obtained in \eqref{Lalgo} and \eqref{mafraF} are identical. However, the appearance of the  unphysical poles in the double-cover framework, $(P_{\!5:2}^2\!-\!P^2_{\!12})^{-1}=(P_{\!34}^2\!-\!P^2_{\!3:5})^{-1}$ and $(P^2_{\!3:5}\! -\!P^2_{\!34})^{-1} $ makes it clear that the two representations are not directly equal. 
Interestingly,  these unphysical poles are related to the physical channel, $\frac{1}{P_{\!3:5}^2\, P_{\!34}^2} $ by use of the partial fraction identity
\begin{equation}
\frac{1}{P_{\!3:5}^2\, P_{\!34}^2} = \frac{1}{P_{\!3:5}^2\, (P_{\!34}^2 - P_{\!3:5}^2)} + \frac{1}{ P_{\!34}^2\, (P_{\!3:5}^2 -P_{\!34}^2 )} .
\end{equation}
As it happens with the linear propagators at loop-level \cite{Geyer:2015bja,Cardona:2016bpi,Cachazo:2015aol,Baadsgaard:2015hia}, the
CHY-formalism is naturally built of linear propagators that can relate to the usual Feynman propagators by means of partial fractioning.

\section{Conclusions} 

We have presented a new set of factorization identities for Yang-Mills theory that naturally
arise from a double-cover version of the CHY-formalism. These factorizations glue amplitudes
together in what can interpreted as the covariant Feynman gauge, with the additional 4-point
contact interactions coming from an explicit sum over longitudinal polarizations. 
The factorizations are at the conjectured level, but there are many
hints that they may also derivable from Berends-Giele recursions. Although spurious poles
appear, simple checks show that they cancel through repeated use of partial fraction identities.
It would be an interesting extension of this work to derive these relations directly from off-shell recursion relations.

Factorizations of amplitudes grow out of the double-cover formalism precisely because it is
``double'': there are, figuratively speaking, two CHY-integrals involved. The bridge between
these two CHY-integrals is an off-shell leg, a propagator. In the double-cover formalism this
off-shell leg stems from one scattering equation that is not imposed as a delta-function constraint.

These factorizations of Yang-Mills amplitudes are just a small part of more general relations
that follow when the double-cover formalism of CHY is analyzed for the known set of theories
that can represented in this form. Details of that will be provided by one of us in a subsequent
paper \cite{Gomez:2018cqg}.

\vspace{0.2cm}
\begin{acknowledgements}
{\sc Acknowledgements:} ~Numerous discussions with J. Bourjaily are gratefully acknowledged. We also thank
N. Ahmadiniaz, C. Cardona, and C. Lopez-Arcos for useful discussions and comments. 
This work was supported in part by the Danish National Research Foundation (DNRF91).
\end{acknowledgements}




\end{document}